\documentclass[letterpaper]{article}
\usepackage[margin=1in]{geometry}
\usepackage[utf8]{inputenc}
\usepackage[shortlabels]{enumitem}
\usepackage{amsmath, mathtools, amssymb, fancyhdr, lastpage, graphicx, epstopdf, listings, hyperref, caption, subcaption, float, array, latexsym, amsthm, graphics, amsfonts, setspace, esvect, color, pdfpages, hyperref, titling}

\usepackage{scalerel,stackengine}
\stackMath
\newcommand\reallywidehat[1]{%
\savestack{\tmpbox}{\stretchto{%
  \scaleto{%
    \scalerel*[\widthof{\ensuremath{#1}}]{\kern-.6pt\bigwedge\kern-.6pt}%
    {\rule[-\textheight/2]{1ex}{\textheight}}
  }{\textheight}%
}{0.5ex}}%
\stackon[1pt]{#1}{\tmpbox}%
}
\hypersetup{
    colorlinks = true,
    linkcolor = black,
    citecolor = blue
}

\usepackage{color} 
\definecolor{mygreen}{RGB}{34,139,34} 
\definecolor{mylilac}{RGB}{160,32,240}


\newcommand{\bit}{\begin{itemize}}
\newcommand{\eit}{\end{itemize}}
\newcommand{\ben}{\begin{enumerate}}
\newcommand{\een}{\end{enumerate}}
\newcommand{\sben}{\begin{shortenumerate}}
\newcommand{\seen}{\end{shortenumerate}}
\newcommand{\bds}{\begin{description}}
\newcommand{\eds}{\end{description}}

\newcommand{\mbf}{\mathbf}

\usepackage{amsfonts}
\usepackage{amsmath}
\usepackage{amssymb}
\usepackage{amsthm}
\usepackage{graphics}
\usepackage{graphicx}
\usepackage{url}
\usepackage{multicol}
\usepackage{float}
\usepackage{ifthen}
\usepackage[leftcaption]{sidecap}
\sidecaptionvpos{figure}{c}
\usepackage[labelfont=bf]{caption}
\usepackage[nocompress]{cite}

\usepackage{csquotes}
\usepackage[english]{babel}

\usepackage[T1]{fontenc}

\usepackage{color} 
\definecolor{mygreen}{RGB}{34,139,34} 
\definecolor{mylilac}{RGB}{160,32,240}

\usepackage{graphicx}
\graphicspath{ {./Images/} }
\usepackage{wrapfig}

\usepackage{makecell}

\title{Quantification of Flagellar Gait Changes with Combined Shape Mode Analysis and Swimming Simulations}

\author{Kelli E. Gutierrez$^{1}$, Becca Thomases$^{2}$, Paulo E. Arratia$^{3}$, and Robert D. Guy$^{1}$}

\usepackage{outlines}
\usepackage{enumitem}
\usepackage{mathtools}
\usepackage{MnSymbol}
\setenumerate[1]{label=\Roman*.}
\setenumerate[2]{label=\Alph*.}
\setenumerate[3]{label=\roman*.}
\setenumerate[4]{label=\alph*.}
\usepackage[title]{appendix}

\begin{document}
\renewcommand\maketitlehooka{\null\mbox{}\vfill}
\renewcommand\maketitlehookd{\vfill\null}
\maketitle 
{\noindent \footnotesize $^{1}$ Department of Mathematics, University of California Davis, Davis, CA 95616, USA\\
$^{2}$ Department of Mathematical Sciences,
Smith College, Northampton, MA, 01063, USA\\
$^{3}$ Department of Mechanical Engineering and Applied Mechanics, The University of Pennsylvania,
Philadelphia, PA 19104, USA \\
The work of KEG was partially supported by NSF Graduate Fellowship 2036201.}
\clearpage

\begin{abstract}
Many different microswimmers propel themselves using flagella that beat periodically. The shape of the flagellar beat and swimming speed have been observed to change with fluid rheology. We quantify changes in the flagellar waveforms of \textit{Chlamydomonas reinhardtii} in response to changes in fluid viscosity using (1) shape mode analysis and (2) a full swimmer simulation to analyze how shape changes affect the swimming speed and to explore the dimensionality of the shape space. By decomposing the gait into the time-independent mean shape and the time-varying stroke, we find that the flagellar mean shape substantially changes in response to viscosity, while the changes in the time-varying stroke are more subtle. Using the swimmer simulation, we quantify how the swimming speed is affected by the dimensionality of the flagellar shape reconstruction, and we show that the observed change in swimming speed with viscosity is explained by the variations in mean flagellar shape and beat frequency, while the changes in swimming speed from the different time-varying strokes are on the scale of variation between cells.  
\end{abstract}

\section{Introduction}

Many microorganisms swim in or transport fluids using flagella, which are thin, threadlike filaments that move in a wavelike motion and enable cell swimming in low Reynolds number fluids \cite{Lauga_Powers_2009}. Examples of these microorganisms include sperm \cite{guasto} and small algae species \cite{li}. A eukaryotic axoneme is composed of a characteristic 9+2 structure, which consists of nine microtubule doublets around the cylindrical perimeter of the flagellum and two microtubules in the center \cite{camalet}. When ATP is present, dynein motors drive the sliding of these doublets and generate forces \cite{Lindemann_1994}, which cause flagellar bending and result in a periodic beat. 

The general flagellar structure and the physical mechanisms that govern force generation and bending are known. The beat emerges from the coupled dynamics of the external fluid, the flagellum, and internal motors which are affected by the deformation, but the mechanism of mechanochemical feedback on motor activity is not well understood \cite{bayly2014, hines-blum, lindemann, riedel-kruse}. Evidence of such feedback is evident from the  observed flagellar gait changes in response to fluid rheologies. For example, sperm exhibit substantial changes in flagellar waveform in fluids with viscosity ranging between 150-140,000 times that of water \cite{Gaffney, sperm_ishimoto_gadelha, Woolley_Vernon_2002}. We want to quantify how varying fluid viscosity leads to systematic changes in shape, which is difficult with large ranges of viscosity. Few studies explore changes in flagellar gait with small variations in fluid viscosity; notably, Yagi et al. \cite{yagi}, Qin et al. \cite{qin}, and Wilson et al. \cite{Wilson_Gonzalez_Dutcher_Bayly_2015} document changes in the flagellar beat of \textit{Chlamydomonas reinhardtii} over a narrow viscosity range. We examine experimental data from Qin et al. \cite{qin} over a small range of fluid viscosity, and perform a quantitative analysis of the resulting changes in flagellar gait.

To analyze these flagellar shape changes with viscosity, we compare gaits using shape mode analysis. In 2014, Werner et al. \cite{werner} described how principal component analysis is used to analyze shape change by looking at the flagellar waveforms of sperm in a low dimensional shape space. Shape mode analysis has been used for quantitative comparisons, such as comparing sperm waveforms across populations \cite{guasto, walker} and comparing different microorganisms' gaits under changes in fluid rheology \cite{sperm_ishimoto_gadelha} and calcium levels \cite{Gholami}. In this manuscript, we couple the results of shape mode analysis with a swimming simulation to understand the effect of flagellar shape changes on swimming behavior.

\begin{figure}[ht!]
    \centering
    \includegraphics[width=\linewidth]{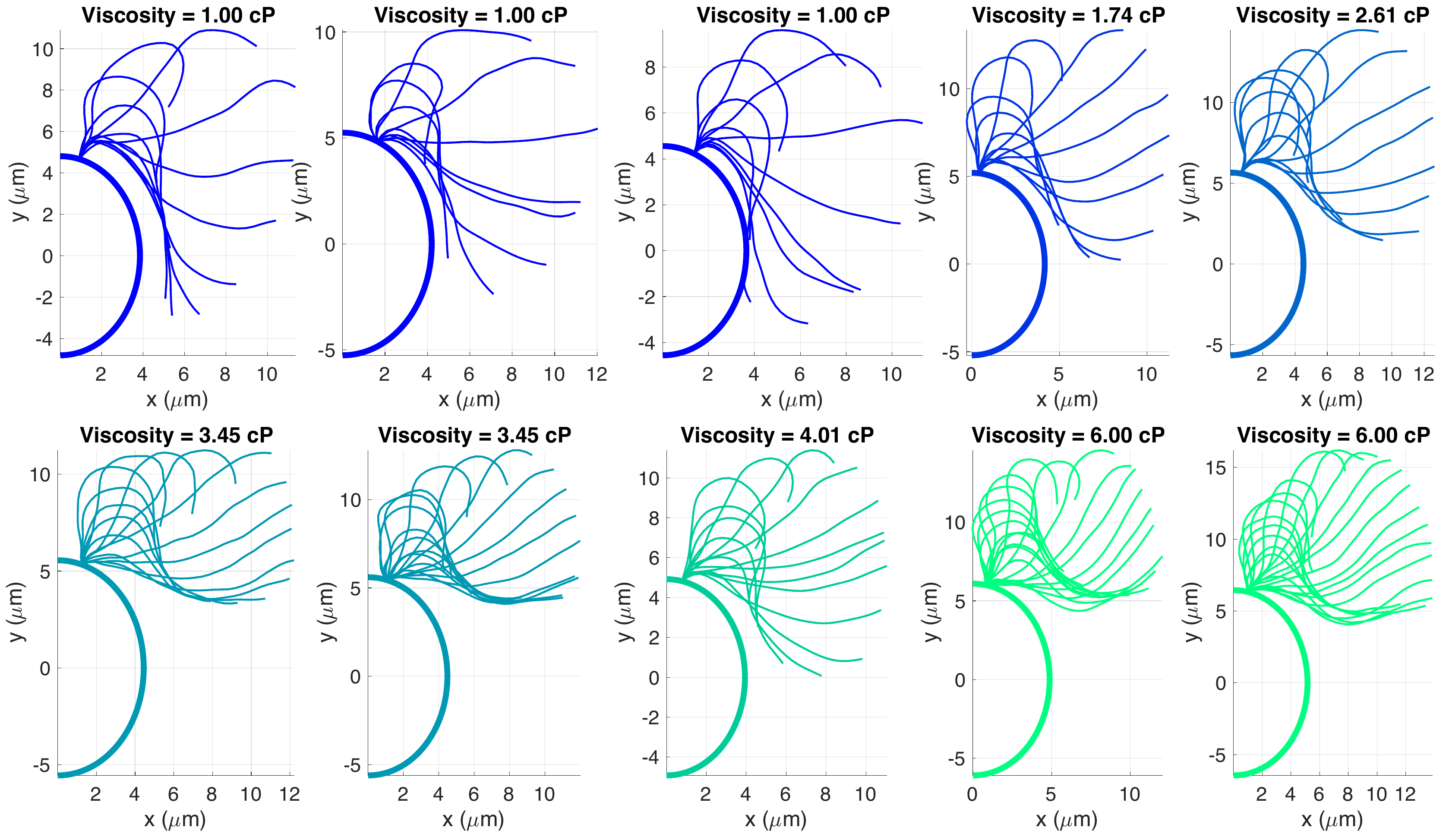}
    \caption{Flagellar stroke patterns over a single period, pictured at time intervals of 0.0017 seconds, attached to a \textit{C. reinhardtii} half cell outline centered at the origin. The body size and the length of the flagellum varies from cell to cell. Each panel corresponds to experimental data from \cite{qin} with individual \textit{C. reinhardtii} cells in Newtonian fluid, with viscosities ranging between 1-6 cP. Fluids with different viscosities are prepared by adding Ficoll (Sigma Aldrich), 5-20\% by weight, to M1 buffer solutions. Experimental details can be found in Appendix \ref{appendix:experiment} and in \cite{qin}.}
    \label{fig:data_strokes}
\end{figure}

\begin{figure}[b!]
    \centering
    \includegraphics[width=\linewidth]{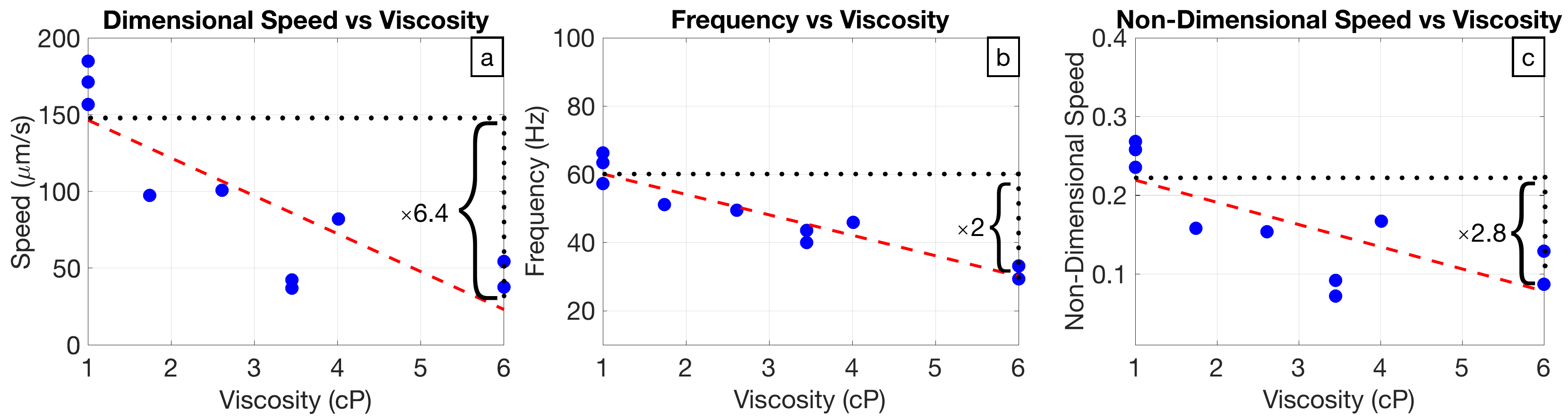}
    \caption{Effect of increasing viscosity from 1.00 cP to 6.00 cP on the swimming speed and the beat frequency. (a) The dimensional speed decreases by approximately a factor of 6.4. (b) The beat frequency decreases by approximately a factor of 2. (c) The non-dimensional swimming speed decreases by approximately a factor of 2.8.}
    \label{fig:data_freq_speed}
\end{figure}

We study flagellar shape changes in \textit{Chlamydomonas reinhardtii} (\textit{C. reinhardtii}) when swimming in fluids with varying viscosity. \textit{C. reinhardtii} are single-celled alga with small bodies about 10 $\mu$m wide that swim with two thin flagella in a breaststroke-like motion to pull themselves forward in a fluid. It is often used as a model organism in cell and molecular biological research to answer key questions, such as how cells generate regular flagellar waveforms \cite{goldstein}. To study flagellar shape changes, we use experimental data \cite{qin} for ten different \textit{C. reinhardtii} cells swimming in Newtonian fluids with viscosities ranging between 1 to 6 cP. A summary of the experimental methods is provided in the Appendix \ref{appendix:experiment}; for more details see Qin et al. \cite{qin}. The data includes the position of the flagellum over the course of several beats, the average swimming speed, and the frequency of the flagellar beat. The strokes over a single period for all of the cells are visualized in Figure \ref{fig:data_strokes} and the frequency of the flagellar beat and the swimming speed of the cell are pictured in Figure \ref{fig:data_freq_speed}. As previously observed \cite{li, Minoura_Kamiya_1995, qin}, the swimming speed and beat frequency both decrease with viscosity, but the speed decreases faster than the frequency, and thus, the frequency change is not entirely responsible for the speed decreasing with viscosity. Figure \ref{fig:data_freq_speed}c shows speed, nondimensionalized by flagellar length and period, decreases with viscosity, which implies that the speed decrease is related to the flagellar gait change.

Shape changes with viscosity have previously been documented for isolated, demembranated, and reactivated axonemes of \textit{C. reinhardtii} \cite{Geyer2022} and for uniflagellar \textit{C. reinhardtii} mutants \cite{Wilson_Gonzalez_Dutcher_Bayly_2015}. By contrast, we analyze flagellar shape changes for wildtype, biflagellated swimming cells. Furthermore, we couple shape mode analysis with a swimming simulation in order to compare experimental data with simulation results and to quantify the effect that flagellar gait changes have on swimming speed.

We first perform shape mode analysis \cite{werner} on the data \cite{qin} in Section \ref{sec:shape_analysis}. Then we explore how the choice of dimension of shape space affects the reconstruction of the flagellar beat and how the flagellar mean shape changes with viscosity. In Section \ref{sec:sim}, we use a swimming simulation based on different shape reconstructions to quantify how changes in mean shape and time-dependent stroke separately contribute to changes in microorganism swimming.

\section{Shape Analysis}\label{sec:shape_analysis}

\subsection{Methods}

We utilize shape mode analysis \cite{werner}, which analyzes different flagellar movement patterns using principal component analysis (PCA) and projects higher dimensional data onto a low dimensional set. We investigate the PCA decomposition of our data to quantitatively compare gaits across fluid rheologies. 

When freely swimming, the flagellar gait of \textit{C. reinhardtii} is generally three-dimensional, particularly when turning or rotating \cite{RufferNultsch, CorteseWan}. The cells in our experiments were confined to a thin film to ensure planar swimming, where the film thickness is about twice the alga body diameter so the cells are unable to rotate about their swimming axis and only swimming in the mid-plane of the film was recorded \cite{qin}. Experimental data is obtained for two-dimensional planar flagellar beating only; data were discarded when the cell body rotated significantly and/or the flagella length deviated by more than 10\%. See Appendix \ref{appendix:experiment} and Qin et al. \cite{qin} for more details.

For each cell, the data consists of the flagellum location in the body frame at time intervals $\Delta t = 0.0017$ seconds for approximately 4-7 full cycles of the periodic beat. At each time point, the positional data consists of 31 approximately equally spaced positions $(x,y)$ along the flagellum. For each point in time, where $t_m = m \Delta t$, we compute the tangent angles at each point and describe the flagellar shape by a vector of tangent angles, $\vec{\psi}(t_m)$. We collect all of these vectors $\vec{\psi}(t_m)$ into an array $\psi_j$ for each cell, where the index $j$ refers to which cell the array describes. The representation of the flagellar shape can be decomposed into two parts: a time-independent mean shape, $\overline{\psi_j}$, and a time-varying stroke. To perform shape mode analysis on the data, we combine all of the demeaned arrays $\psi_j - \overline{\psi_j}$ into a single array:
$$ \Psi = \begin{bmatrix} \psi_1 - \overline{\psi_1}  \hspace{.1in} \vline  \hspace{.1in}  \psi_2 - \overline{\psi_2} \hspace{.1in} \vline  \hspace{.1in}  \hdots  \hspace{.1in} \vline  \hspace{.1in} \psi_j - \overline{\psi_j}  \hspace{.1in} \vline   \hspace{.1in} \hdots \end{bmatrix}.$$
The eigenvectors of covariance matrix $C = \Psi^T \Psi$ define an orthonormal basis of shapes called shape modes \cite{werner}. The shape of the flagellum for the $j$th cell at the time $t_m$ is decomposed as:

\begin{equation} \label{eq:basis}
    \vec{\psi}_j(t_m) = \overline{\psi_j} + \sum_{k} v_k^j(t_m) \mbf{S}_k,
\end{equation} 
where $\mbf{S}_k$ is the $k$th shape mode. The coefficients  $v_k^j(t_m)$, are defined as 
\begin{equation} \label{eq:time_coeff}
    v_k^j(t_m) = \langle \vec{\psi}_j(t_m) - \overline{\psi_j}, \mbf{S}_k \rangle.
\end{equation} 
These time-dependent coefficients characterize how much $\mbf{S_k}$ contributes to the data at a particular point in time.

Using shape mode analysis, we investigate the flagellar shapes and how the shape changes with viscosity. Particularly, we look at the base shapes $\mbf{S}_k$ that make up flagellar gait, investigate low dimensional Fourier series representations of the time-dependent coefficients $v_k^j(t_m)$, and examine how the temporal parts of the stroke and the mean shape change with viscosity.

\subsection{Results} \label{sec:shape_analysis_results}

\subsubsection{Shape Basis Modes}

Eigenvalue decomposition of the covariance matrix produces an orthogonal basis for the shape space. We order the eigenvalues from largest to smallest, which gives an ordering that corresponds to the shape modes' relative contribution to the overall shape of the flagellum. Note this ordering corresponds to that from singular value decomposition of $\Psi$, and therefore the eigenvalues $\lambda_i$ correspond to the singular values $\sigma_i$, i.e.\ $\lambda_i = \sigma_i^2$. 

\begin{figure}[ht]
    \centering
    \includegraphics[width=\linewidth]{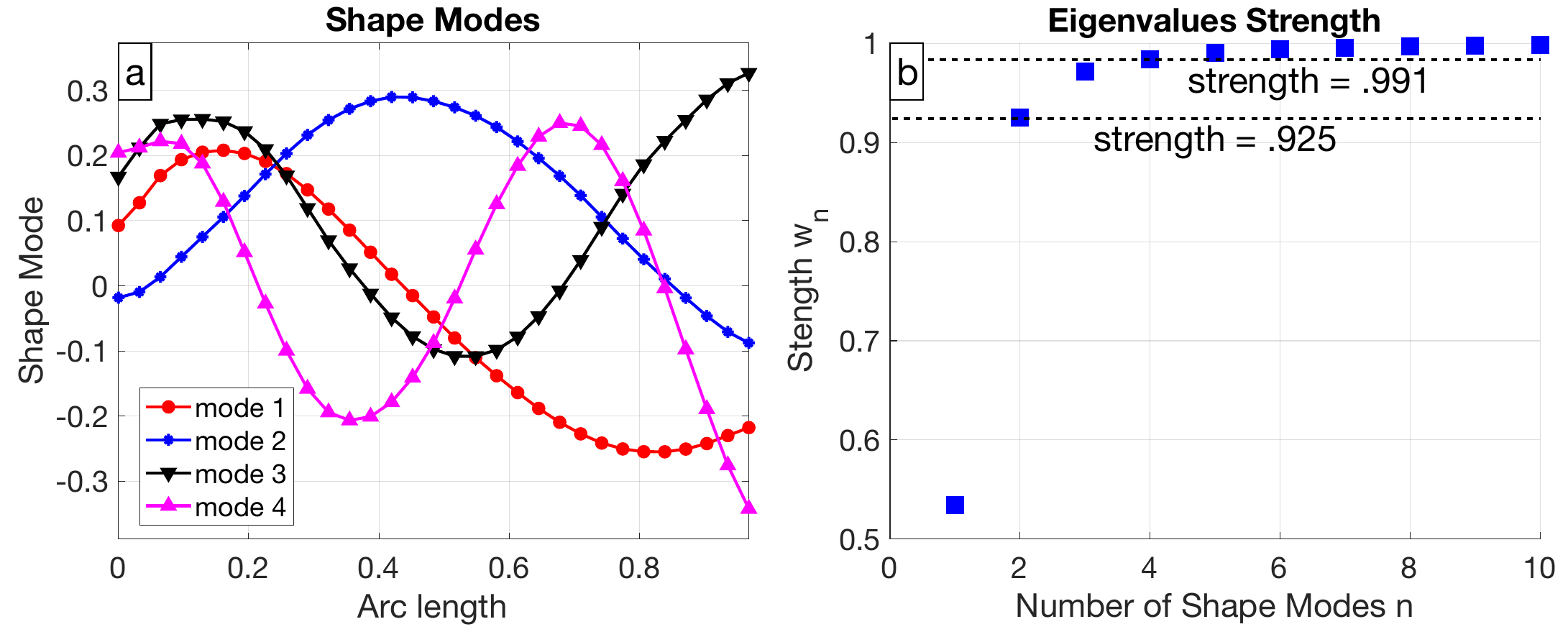}
    \caption{(a) The first four shape modes plotted against arc length, which are the first four eigenvectors of $\Psi^T \Psi$.} (b) The eigenvalue strength $w_n$, defined in equation \eqref{eq:eval_strength}, quantifies the amount of variability captured by $n$ shape modes.
    \label{fig:shapemode}
\end{figure}

The first four shape modes are pictured in Figure \ref{fig:shapemode}a and look similar to shape modes for asymmetric beats such as in \textit{C. reinhardtii} \cite{Gholami} or cilia \cite{saintillian}. The first mode has an "S" shape, the second mode has an "U" shape, and the for higher number shape modes, the number of extrema increases, which is similar to the shape modes for different types of sperm \cite{saintillian, guasto, sperm_ishimoto_gadelha, ishimoto2, nandagiri, oriola, walker, werner}.

We quantify how the amount of variability from the space spanned by the first $n$ shape modes contributes to the total variance in the data \cite{Mackiewicz_Ratajczak_1993}. As shown in Figure \ref{fig:shapemode}b, the eigenvalue strength $w_n$ is defined as the cumulative sum of the eigenvalues divided by the total sum, i.e.
\begin{equation} \label{eq:eval_strength}
    w_n = \frac{\sum_{i=1}^n \lambda_i}{\sum_i \lambda_i}.
\end{equation}
The first two shape modes capture 92.5\% of the data, meaning that only two shape modes are needed to reconstruct the flagellar shapes and include over 90\% of variability. From a statistical point of view, two modes are sufficient to portray the majority of the information about the stroke, which implies the shape space is two dimensional. It is common for studies utilizing shape mode analysis on sperm and \textit{C. reinhardtii} to only use two modes because two modes often capture over 90\% of the variance \cite{saintillian,ishimoto2,nandagiri}. Note that Gholami et al. \cite{Gholami} uses four modes in a \textit{C. reinhardtii} flagellar shape reconstruction, despite two modes capturing over 90\% of the variance. It remains to be seen how data variation from higher indexed shape modes affects the behavior and the swimming speed, which we will explore in Section \ref{sec:sim}.

\subsubsection{Time Dependent Stroke Analysis}

\begin{figure}[ht!]
    \centering
    \includegraphics[width=\linewidth]{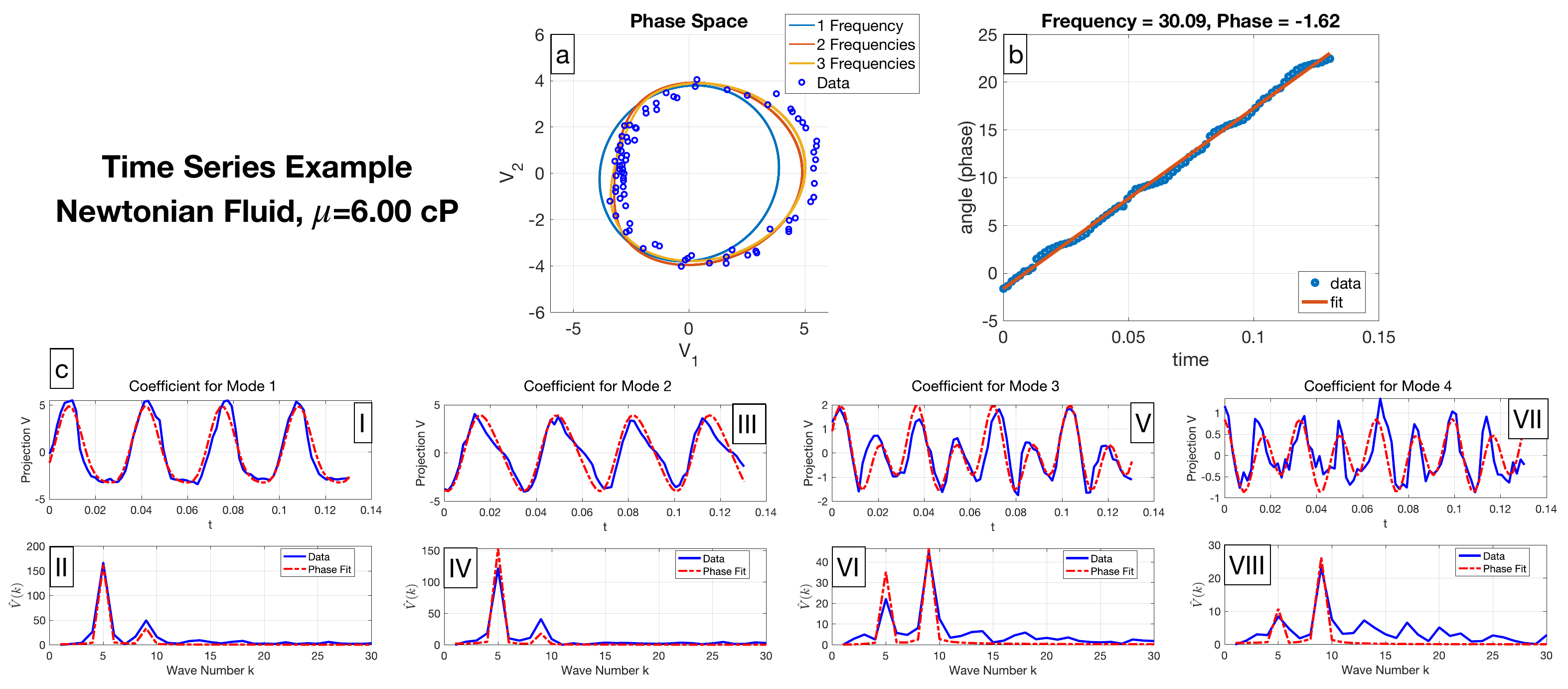}
    \caption{A time series example for one of the cells in a Newtonian fluid with viscosity 6.00 cP. (a) Phase space, where the blue data points represent the projection of the experimental data onto the first two shape modes, according to Equation \eqref{eq:time_coeff}, and the Fourier series representations for $v_1$ and $v_2$ with different frequencies are visualized with the different colored loops. (b) Linear fit for the polar angle in the $v_1$-$v_2$ plane, where the slope is the angular frequency and the $y$-intercept is the initial phase. (c) Fourier series representations for $v_1-v_4$, with plots (I,III,V,VI) in real space against non-dimensional time and plots (II, IV, VI, VIII) in Fourier space against non-dimensional frequency.}
    \label{fig:time_series}
\end{figure}

The shape modes make up the general shape of the flagellar stroke, and we investigate how these shapes change over a period using the time-dependent coefficients. Recall $v_k^j$ is described as a time-dependent coefficient of the $j$th cell associated with the $k$th shape mode, as defined in Equation \eqref{eq:time_coeff}. We visualize $v_1^j$ and $v_2^j$ in the phase space in Figure \ref{fig:time_series}a. 
Each point corresponds to a flagellar shape at a particular point in time from the data projected onto a two dimensional shape space. The collection of these points from this sample cell are expected to live on a closed curve since the flagellar beat is periodic \cite{Klindt_Ruloff_Wagner_Friedrich_2016, Klindt_Ruloff_Wagner_Friedrich_2017, Ma_Klindt_Riedel-Kruse_Julicher_Friedrich_2014, Solovev_Friedrich_2022, werner} and the data covers several periods of the flagellar beat. The polar angle $\phi$ between $v_1^j$ and $v_2^j$ as a function of time along with its linear fit, $\phi(t) = \omega t + \phi_0$, are shown in Figure \ref{fig:time_series}b. The slope of the line, $\omega$, is the angular beat frequency and the intercept, $\phi_0$, is the initial angle, i.e. the location along the loop where the cell begins in its flagellar beat. Due to the linearity of the data in Figure \ref{fig:time_series}b, the angular speed is constant, meaning 
$$ \frac{d\phi}{dt} = \omega.$$
This angle $\phi$ reasonably approximates phase, so we describe the temporal data as functions of the phase angle. Thus, the time-dependent coefficients $v_k^j(t_m)$ are fit with a Fourier series as a function of the phase angle $\phi(t)$ as follows:
\begin{equation}\label{eqn:v_fit}
    V_{k,N}^j(t_m) = \sum_{n = 1}^N A_{k,n}^j \sin(n\phi(t_m)) + B_{k,n}^j \cos(n\phi(t_m)).
\end{equation}
Figure \ref{fig:time_series}c shows the time-dependent coefficients for the first four shape modes and a Fourier series phase fit with two frequencies for each set of coefficients. From this Fourier series representation, the question that arises is what value of $N$ in equation \eqref{eqn:v_fit} should be chosen to represent the closed loops in Figure \ref{fig:time_series}. 

In order to choose how many frequencies are needed in each Fourier series fit, we start by examining the time-dependent coefficients in Fourier space in Figure \ref{fig:time_series}c. In Figures \ref{fig:time_series}c(II, IV), there is a strong peak at the dominant frequency and the first harmonic has a smaller but significant peak that is approximately a third of the height. Turning back to phase space in Figure \ref{fig:time_series}a, which focuses on the data projected in two dimensional phase space, it is clear that the first harmonic, or second frequency, makes a substantial difference in the fit, but the third frequency does not make a significant visual difference. In Figures \ref{fig:time_series}c(VI, VIII), the dominant peak is at the second frequency, with the first frequency peak being at approximately half of its height. This implies that if the reconstruction has more than two shape modes, at least two frequencies should be used in the Fourier series phase fit. It is important to note that the third and fourth shape modes contribute to 6.6\% of shape space, as is shown in Figure \ref{fig:shapemode}b, so higher frequencies than the dominant one contribute less to the overall stroke. There is no clear way to choose the number of frequencies to use in Fourier series phase fits solely based on Figure \ref{fig:time_series}. We investigate the swimming speed to see how different Fourier series representations affect the flagellar behavior in Section \ref{sec:sim}.

\subsubsection{Phase Space - Comparing All Cells}

\begin{figure}[h]
    \centering
    \includegraphics[width=\linewidth]{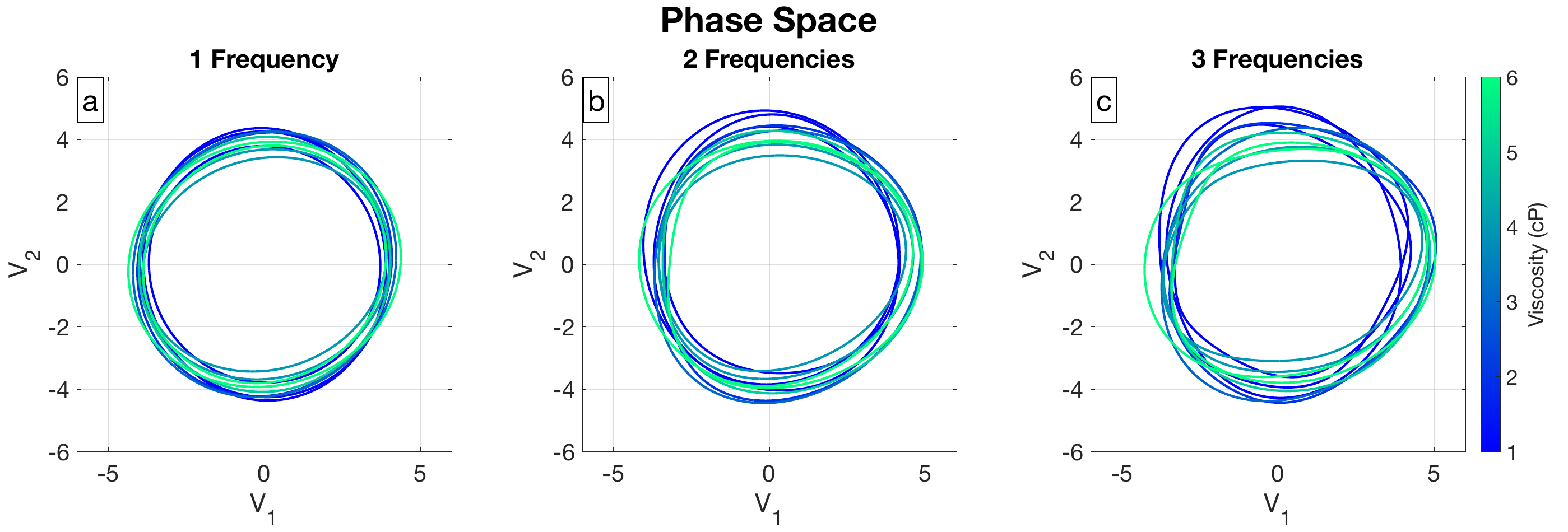}
    \caption{Phase Space Comparison. All the figures show the Fourier series representation from \eqref{eqn:v_fit} of $V_1$ and $V_2$ plotted in phase space, colored according to viscosity, where curves in (a) have one frequency, those in (b) have two frequencies, and those in (c) have three frequencies.}
    \label{fig:phase_space}
\end{figure}

In Figure \ref{fig:phase_space}, we examine the phase space to compare the different number of frequencies used in Fourier series representations from \eqref{eqn:v_fit} for the time-dependent coefficients across all cells. The phase loops that have Fourier series fits with only one frequency are similar in shape, and it is difficult to distinguish between loops of different viscosities. The shapes of the loops for multiple frequencies are generally less elliptical. The loops for lower viscosities are elongated in the $V_2$ direction and the loops for higher viscosities are elongated in the $V_1$ direction, which is visually apparent as the number of frequencies increases. We show in Appendix \ref{appendix:phase_space} that this trend in elongation of the loops is true even with only a single frequency, which is not obvious in Figure \ref{fig:phase_space}a. Since it is not clear how the number of frequencies affects the flagellum shape or the swimming behavior, we further explore how to choose a Fourier series representation for the time-dependent coefficients in Section \ref{sec:sim}.

\subsubsection{Flagellar Mean Shape Changes with Fluid Viscosity}
\begin{figure}[ht!]
    \centering
    \includegraphics[width=\linewidth]{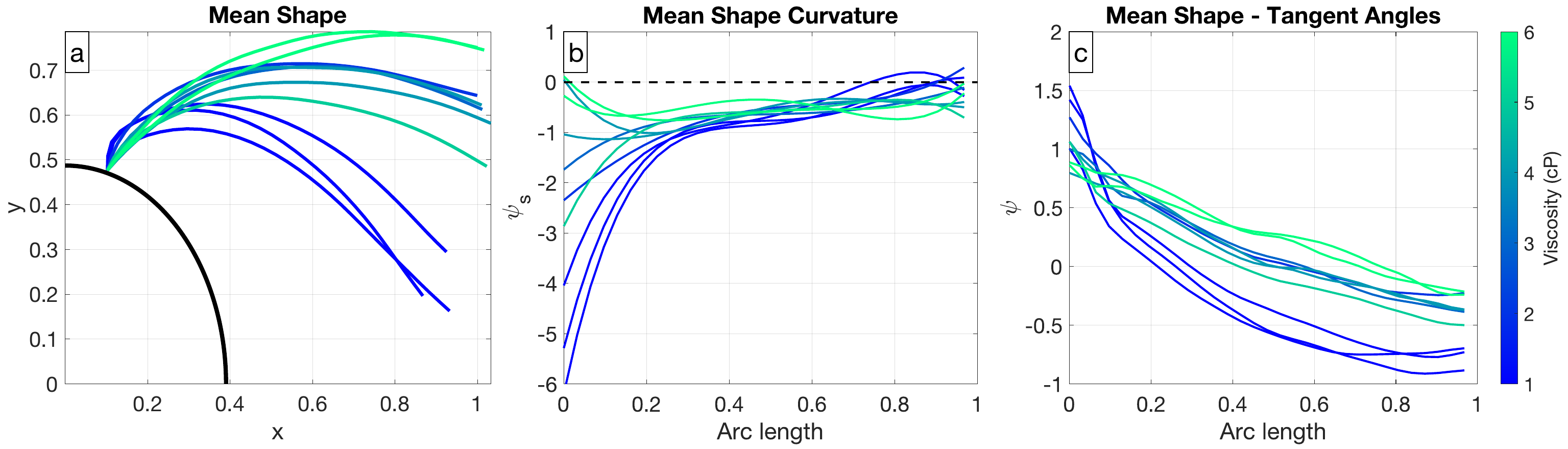}
    \caption{Different representations of mean flagellar shape, colored according to viscosity. (a) Non-dimensionalized time-averaged position of the flagellum, attached at a common point to a partial \textit{C. reinhardtii} cell centered at the origin. (b) Non-dimensional curvature of mean shape, $\psi_s$, plotted against arc length. (c) Tangent angles, $\psi$, along the flagellum of the mean shape. Arc length zero corresponds to the base.}
    \label{fig:mean}
\end{figure}

The previous sections focus on analyzing variations about the mean flagellar shape, and here we narrow our focus on the time-independent mean shape itself. Figure \ref{fig:mean}a shows the time-averaged position of the flagellum corresponding to the mean shape. It is clear that the mean shape changes with viscosity, where at lower viscosities, the time-averaged flagellar position is closer to the body, and at higher viscosities, the time-averaged position is in front of the body. The same visual trend is apparent in Figure \ref{fig:data_strokes}, where the time-dependent flagellar beat is closer to the body at lower viscosities and in front of the body at higher viscosities. The time-independent shape appears more curved for lower viscosities, particularly near the base. In Figure \ref{fig:mean}b, we examine the curvature of the mean shape along the flagellum. The curvature is larger near the basal end in lower viscosity fluids and is relatively constant near the distal end of the flagellum in all fluids.

Previous studies have found fairly constant curvature of the time-independent shape of approximately $ 0.2 \mu m^{-1}$ for isolated and reactivated axonemes of \textit{C. reinhardtii} \cite{GEYER20161098, Gholami_Light} and uniflagellar mutants of \textit{C. reinhardtii} \cite{EshelBrokaw} in buffer fluids with viscosity 1 cP. Examining Figure 9c in Eshel and Brokaw \cite{EshelBrokaw}, which shows the tangent angles along the flagellum of the mean shape, we notice larger curvature close to the attachment point. In Figure \ref{fig:mean}c, we plot tangent angles for swimming, biflagellated \textit{C. reinhardtii} and observe similarly shaped curves near the base at low viscosities. Although we observe non-constant curvature of the mean shape at low viscosities, the spatial average of the curvature at viscosity 1 cP is $\approx .195 \mu m^{-1}$, which aligns with previous studies \cite{GEYER20161098, Gholami_Light}. The spatial average of the curvature can be read from Figure \ref{fig:mean}c using
\begin{equation}\label{eqn:avg_curv}
    \overline{\kappa}_{avg} = \int_0^1 \overline{\kappa} ds = \int_0^1 \overline{\psi_s} ds = \overline{\psi}(1) - \overline{\psi}(0).
\end{equation}
We observe that the spatial average of the curvature of the mean shape decreases with fluid viscosity. Note that flagellar lengths range between $10.28-14.00 \mu m$.

These trends in mean shape with viscosity are more obvious to the eye than the subtle trends and changes in shape seen in previous sections. What is unknown is how these trends for mean shape along with the subtle trends in time-varying stroke affect the organism's swimming speed, which we investigate in Section \ref{sec:sim}.

\section{Swimming Simulations} \label{sec:sim}
As shown in Figure \ref{fig:data_freq_speed}, the swimming speed, the frequency of the flagellar beat, and the non-dimensional speed all decrease with viscosity. This implies the frequency and the flagellar stroke change with viscosity and both directly affect the swimming speed. In this section we quantify how the changes in stroke affect the speed using a swimming simulation.

\begin{figure}[ht!]
    \centering
    \includegraphics[width=.75\linewidth]{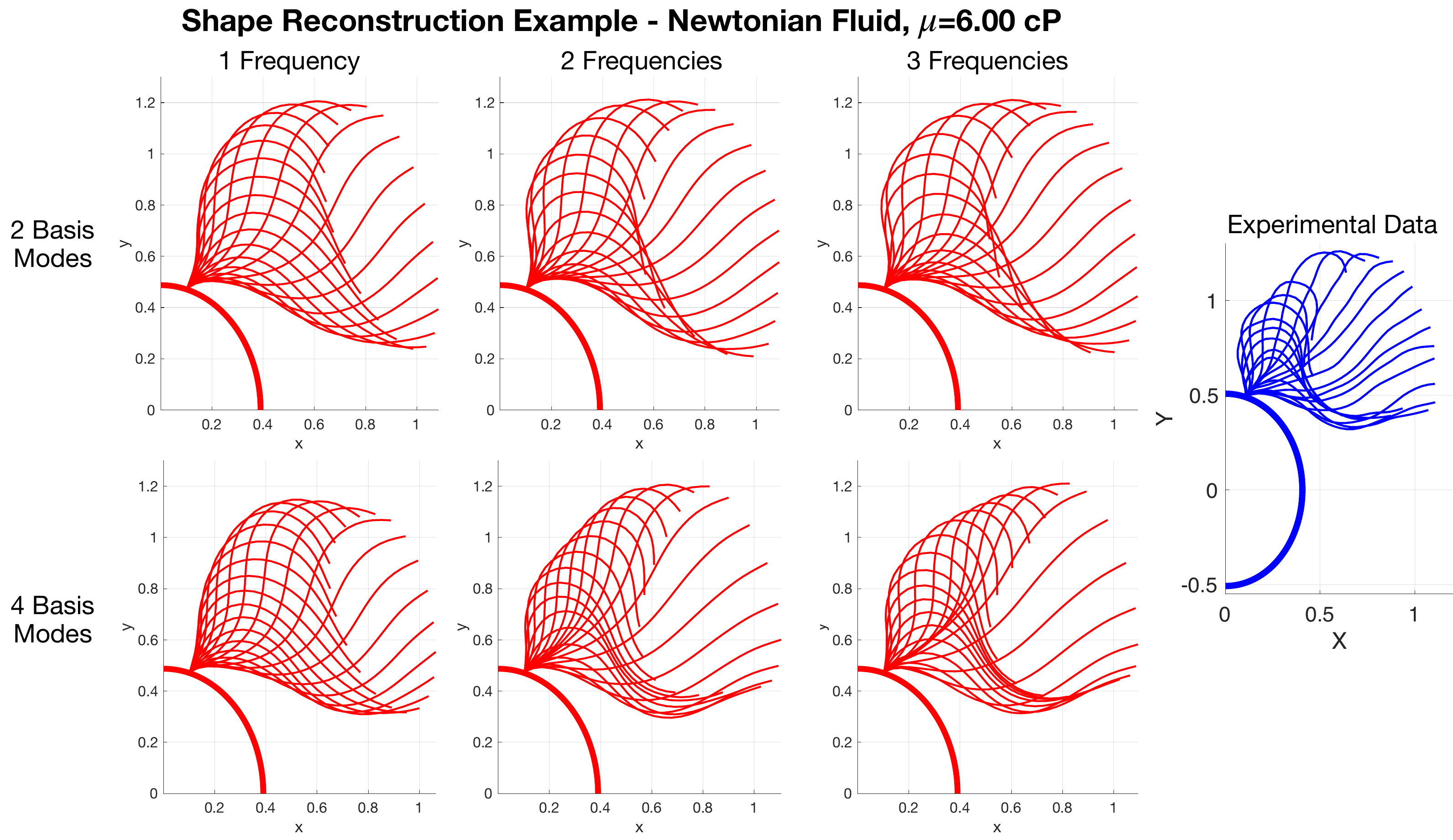}
    \caption{Flagellar shape reconstruction example, attached to a partial \textit{C. reinhardtii} cell centered at the origin, in a Newtonian fluid with viscosity 6.00 cP, using different numbers of shape basis modes and frequencies. The difference between reconstructions and the experimental data is most apparent near the flagellum tip. The length of the flagella is scaled to be one in all panels and the cell body size is uniform.}
    \label{fig:shape_reconstruction}
\end{figure}

Beyond recognizing how change in shape with viscosity affects swimming speed, we want to understand how the choice of low dimensional reconstruction affects the swimming behavior. Figure \ref{fig:shape_reconstruction} shows shape reconstructions for different numbers of shape basis modes and different numbers of frequencies in the Fourier series fit for the time-dependent coefficients. As expected, when the number of shape modes and frequencies increases, the reconstructed stroke approaches the data. Specifically, the behavior near the flagellum tip looks different in each reconstruction. We analyze how each reconstruction affects the speed to determine which shape reconstruction to use in simulations.

\subsection{Methods}
\begin{figure}[hb!]
    \centering
    \includegraphics[width=0.9\linewidth]{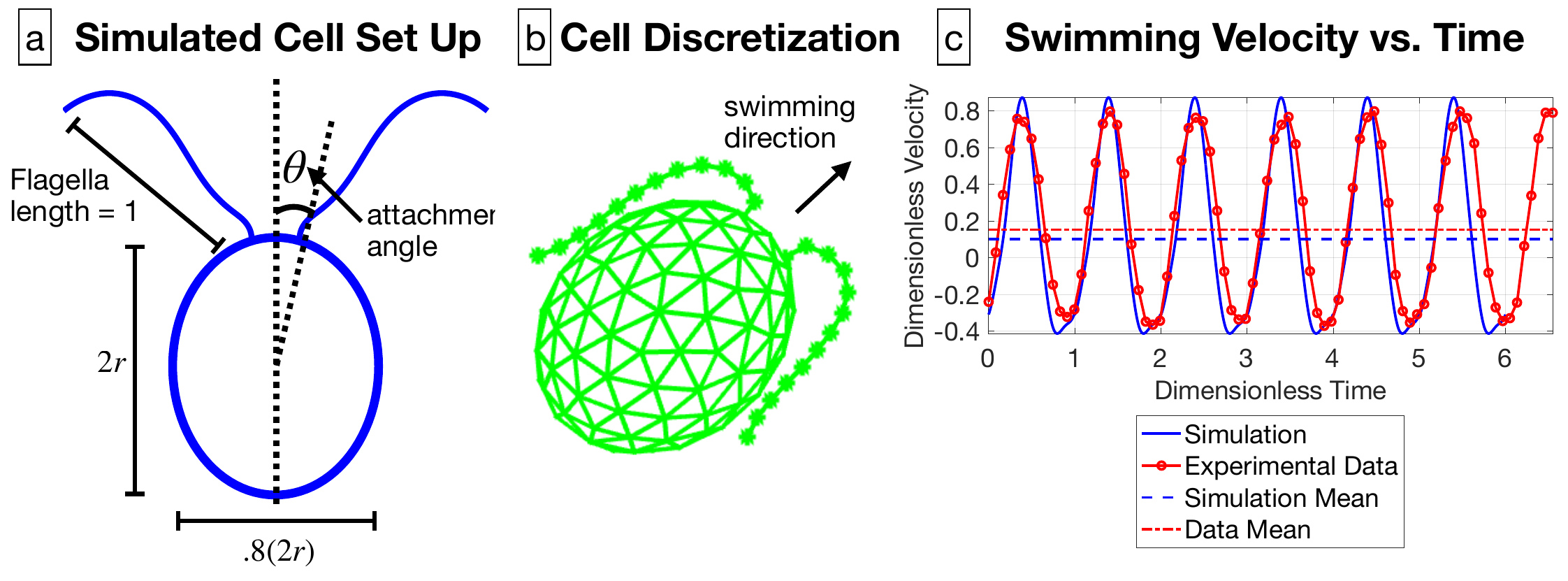}
    \caption{(a) A schematic of the \textit{C. reinhardtii} cell model in 2 dimensions, including the cell body radius $r = 0.4873$, flagellar length, and attachment angle $\theta = 0.2147$ radians. (b) The discretization of the cell body and the flagella. (c) The swimming velocity from the experimental data and the simulation are compared, along with the mean swimming speed, for the cell swimming in viscosity 2.61 cP.}
    \label{fig:simulation_schematic}
\end{figure}

We develop a three-dimensional computational model of a \textit{C. reinhardtii} cell swimming in a Newtonian fluid at zero Reynolds number. An ellipsoidal body is used to model the cell, and the flagella are attached at the angle $ \theta = 0.2147$ radians from the major axis, where this angle is generated by the average of the attachment angles from all cells, as shown in Figure \ref{fig:simulation_schematic}a. The simulation is done non-dimensionally, where the lengths are scaled by the flagellar length and the non-dimensional time scale is the period of the flagellar beat, as found in the linear fit in Figure \ref{fig:time_series}b. Thus, the flagella are given length one. The major radius of the cell body in the swimming direction is $r = 0.4873$, which is chosen as the average across all cells in the data set. The other radii of the body are chosen to be 80\% of the length of the major radius, since \textit{C. reinhardtii} are generally ellipsoidal \cite{Harris_1989}. 

The flagellum is discretized into points separated by distance $\Delta s$, and the ellipsoidal cell body is discretized into points approximately spaced by $\Delta s$, as shown in Figure \ref{fig:simulation_schematic}b. The flagellar gait is prescribed from the shape reconstructions in Section \ref{sec:shape_analysis}, where the position and velocity are given in the body frame. We use the method of regularized Stokeslets \cite{cortez} to solve for the forces on the structure, and the swimming speed is determined based on the constraint that the net force on the swimmer is zero. An example of the swimming velocity from the simulation is pictured in Figure \ref{fig:simulation_schematic}c; see Appendix \ref{appendix:sim_details} for more details.

\subsection{Results} \label{sec:sim_results}
\subsubsection{Reconstruction Shape Space Dimension Determination}
In this section we utilize swimming speeds from the numerical simulation to quantify how the swimming speed is affected by the number of shape basis modes and frequencies used for the flagellar shape reconstruction.

\begin{figure}[htb!]
    \centering
    \includegraphics[width=0.5\linewidth]{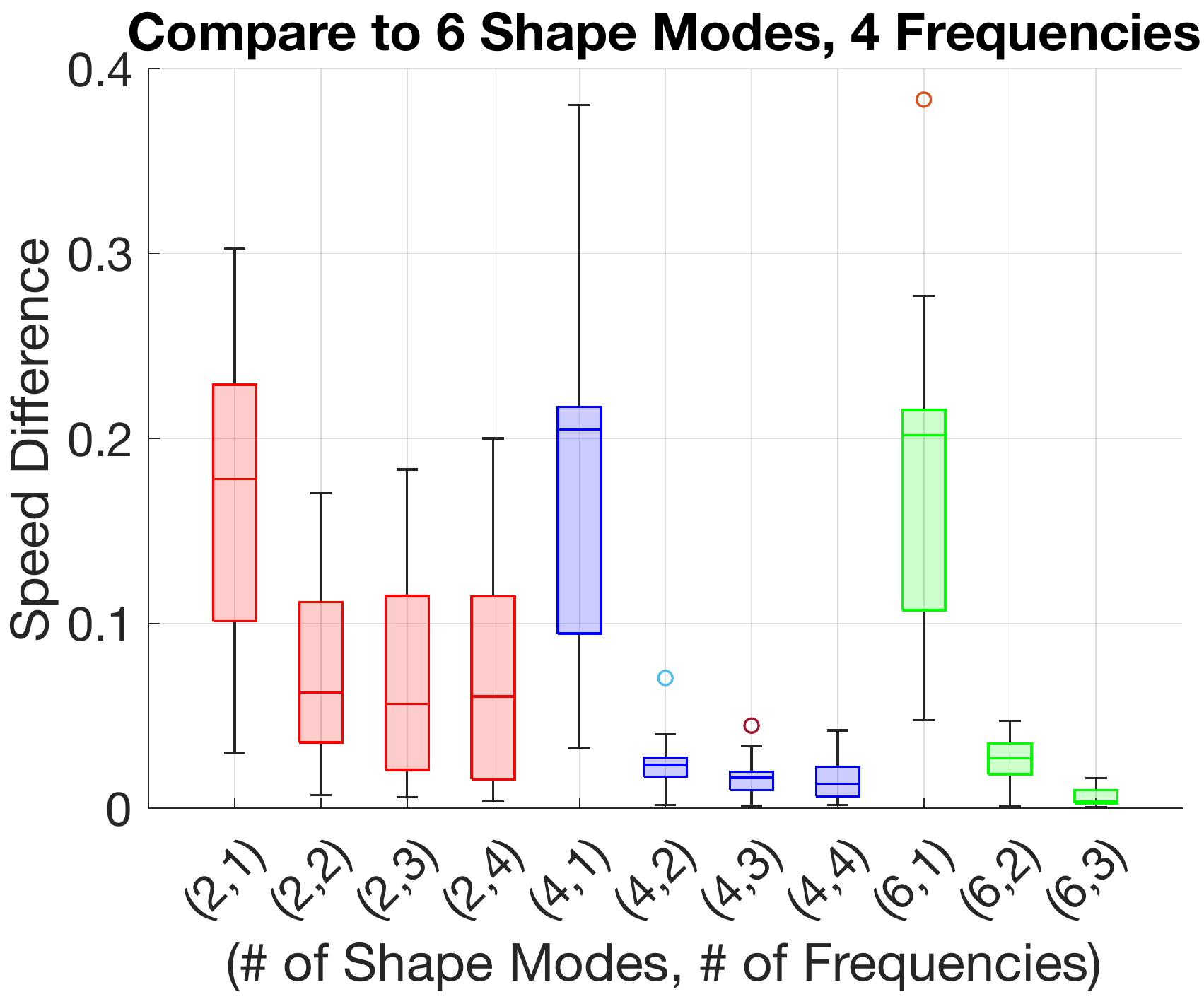}
    \caption{Relative differences in swimming speed, where swimming speeds from simulations with reconstructions using different numbers of shape modes and frequencies are compared to the swimming speed from a simulation using the reconstruction with six shape modes and four Fourier frequencies.}
    \label{fig:velocity_compare}
\end{figure}

We compare different numbers of shape modes and frequencies for the time-dependent coefficients in Figure \ref{fig:velocity_compare}. Based on shape mode analysis, six shape modes statistically gives 99.4\% of the shape data. Generally, for the Fourier series representations of the time-dependent coefficients, as seen in one example in Figure \ref{fig:time_series}, there are two dominant frequencies in most cells, where some cells have notable peaks at the third or fourth frequencies (ie, the second or third harmonic) in higher shape modes. Thus, we compare speeds to those from a reconstruction using six shape modes and four frequencies.

In Figure \ref{fig:velocity_compare}, the relative speed difference for single frequencies is essentially the same for different numbers of shape modes. First, we focus on two shape modes, where the relative speed differences for different number of frequencies are as large as 20\% and average around 5\%, as shown with the red bars in Figure \ref{fig:velocity_compare}. When increasing to four shape modes and two frequencies, the speed difference is generally less than 3\% with an average change close to 2\%. As the number of frequencies and shape modes is increased beyond four shape modes and two frequencies, relative speed differences drop below 2\%. Thus, as shown in Figure \ref{fig:velocity_compare}, even though the results do not vary greatly between reconstructions, we choose to use four shape modes and two frequencies in the reconstruction for the rest of our analysis.

\subsubsection{Speed Decreases with Viscosity}
\begin{figure}[hb!]
    \centering
    \includegraphics[width=.5\linewidth]{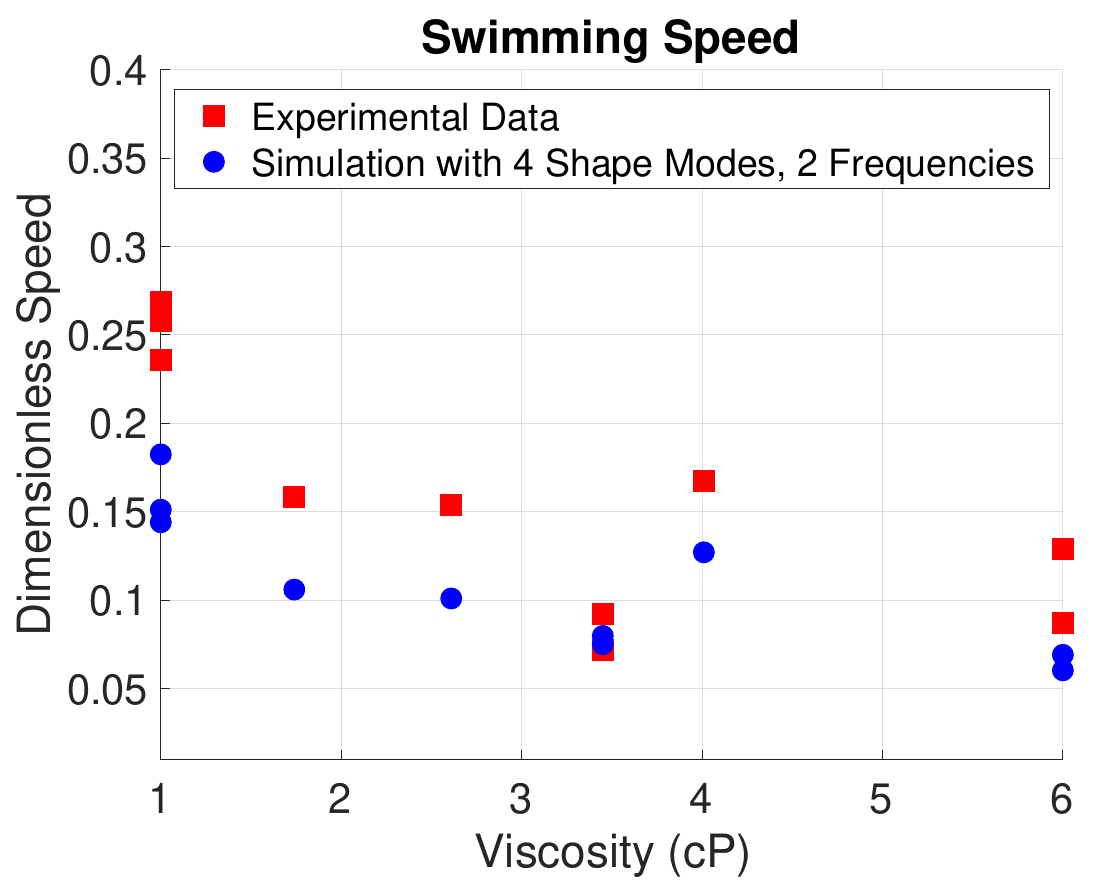}
    \caption{A comparison of experimental and simulated non-dimensional speed for all cells, where the simulation is based on a reconstruction with four shape modes and two Fourier modes.}
    \label{fig:speed_data_model_compare}
\end{figure}

The cell swimming speed for the experimental data and the simulation are compared using the reconstruction with four shape basis modes and two frequencies in Figure \ref{fig:speed_data_model_compare}. There is some variation between the data and the simulation, but many factors may account for these differences. The body size, length of the flagella, and position at which the flagella were attached to the body are all held fixed in the simulation, but these vary slightly in the data due to cell variation. The simulation was computed in free space, but the experiment was performed in a thin film to keep the cells swimming in the plane. These changes in the simulation may account for the data and model differences. Despite these differences, the overall trend of the dimensionless speed from the simulation is the same as that of the data. As the viscosity increases from 1.00 to 6.00 cP, the mean speed is approximately halved in both the simulation and the data. 

\subsubsection{Changes in Mean Shape Primarily Cause Speed to Decrease with Viscosity}

\begin{figure}[ht!]
    \centering
    \includegraphics[width=.72\linewidth]{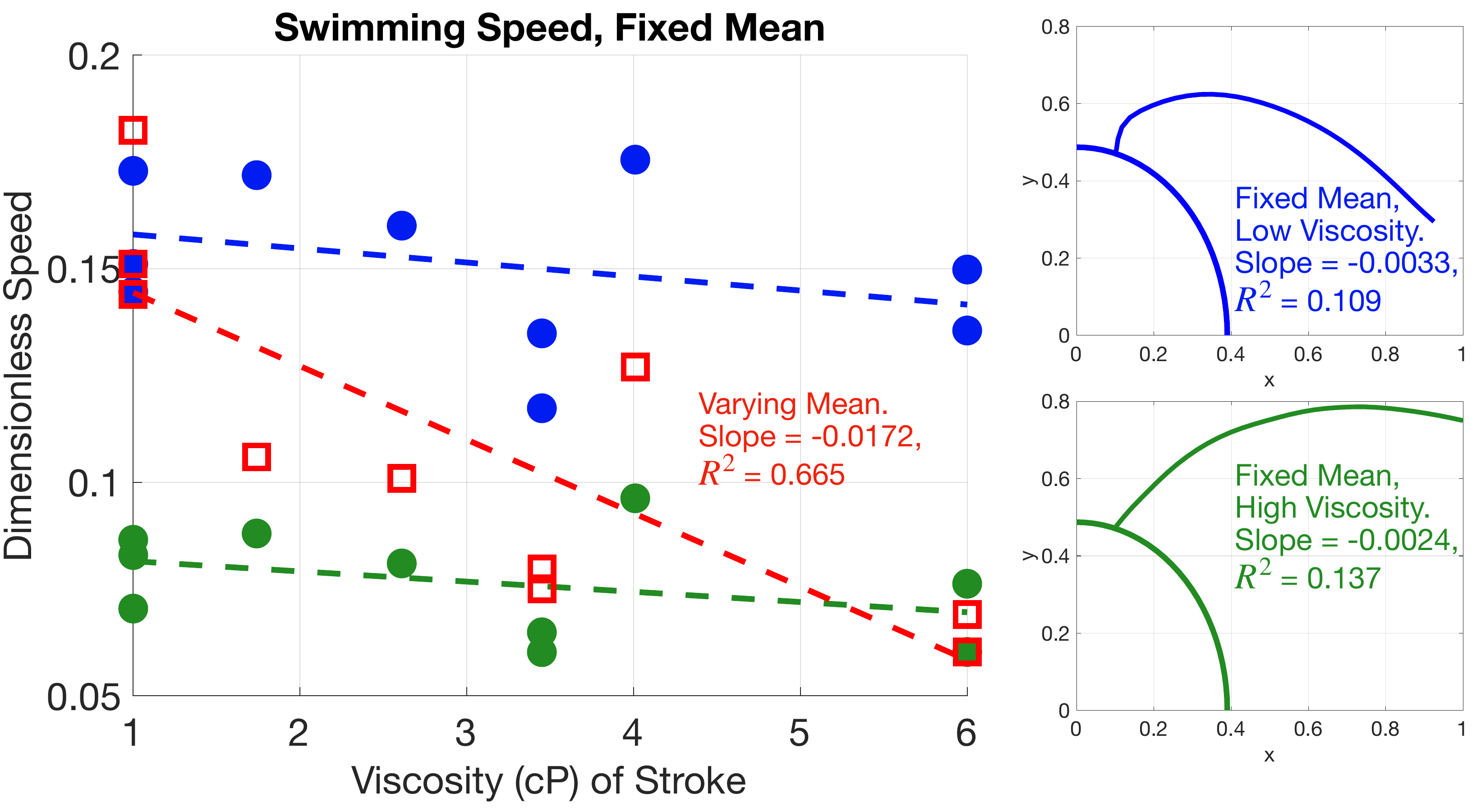}
    \caption{Speed comparison for simulated swimmers where the mean flagellar shape is fixed from cells in fluid with viscosities 1 and 6 cP, and the contributing stroke varies. The horizontal axis refers to the viscosity of the fluid from the cell which contributes the time-varying stroke. Blue circles give speeds from the simulation with a fixed mean shape from a cell in viscosity 1.00 cP and green circles give speeds from the simulation with a fixed mean shape from a cell in viscosity 6.00 cP, where mean shapes attached to a partial \textit{C. reinhardtii} cell are shown on the right of the figure in the corresponding color. The open red squares depict the speed where the mean shape and stroke both vary with viscosity and are the same as the blue dots in Figure \ref{fig:speed_data_model_compare}.}
    \label{fig:fixed_mean}
\end{figure}

In Figure \ref{fig:speed_data_model_compare}, the mean shape and the time-varying stroke from each cell change with viscosity and the non-dimensional speed decreases with viscosity. It is not clear how the changes in mean shape and stroke contribute to this change in speed, and thus we vary these two factors independently to quantify how each affects the swimming speed.

\begin{figure}[ht!]
    \centering
    \includegraphics[width=0.72\linewidth]{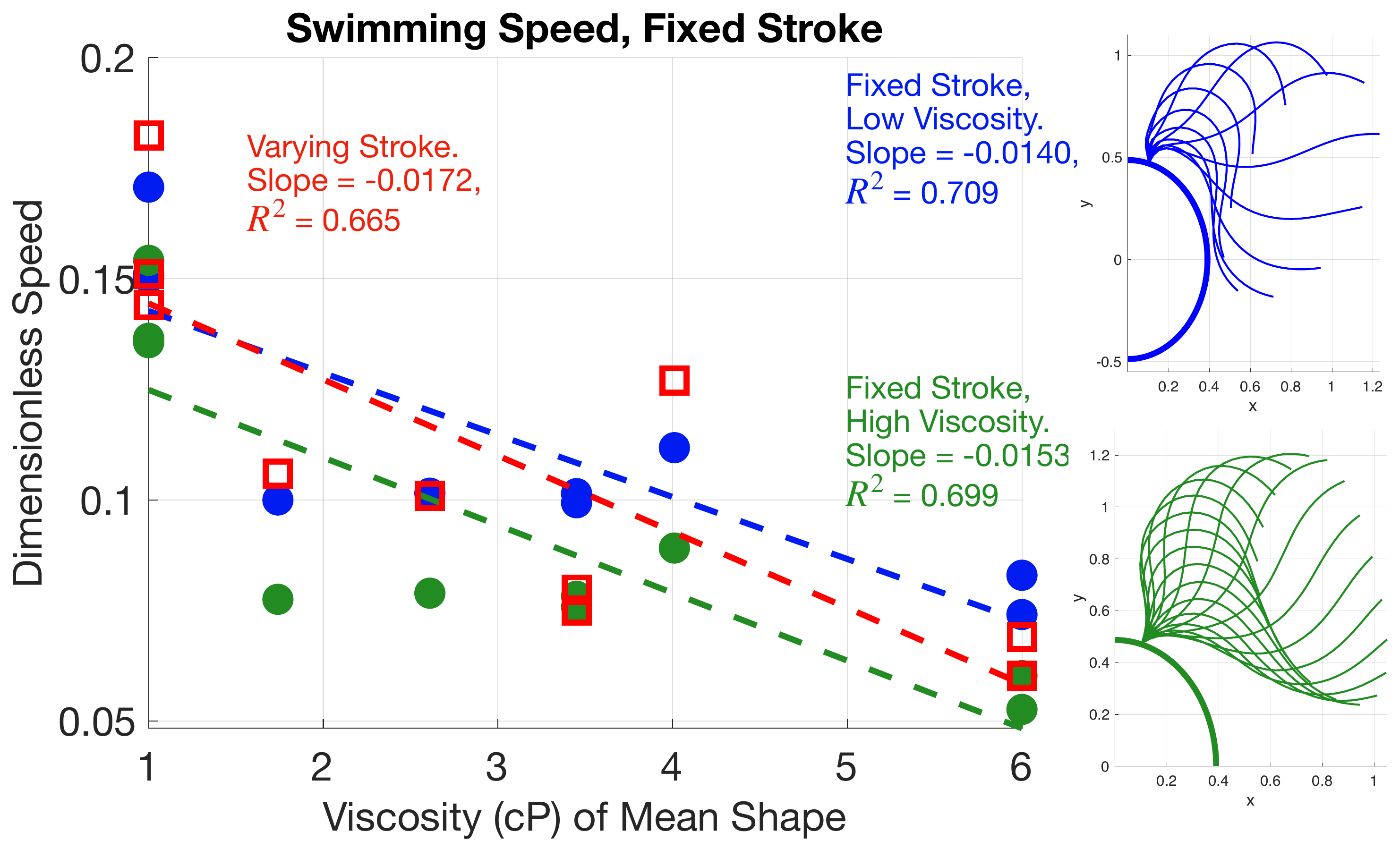}
    \caption{Speed comparison for simulated swimmers where the time-varying stroke is fixed from cells in fluid with viscosities 1 and 6 cP, and the mean flagellar shape varies. The horizontal axis refers to the viscosity of the fluid from the cell which contributes the mean shape. Blue circles represent speeds from the simulation with a fixed stroke from a cell in viscosity 1.00 cP and green circles represent speeds from the simulation with a fixed stroke from a cell in viscosity 6.00 cP, where the strokes attached to a partial \textit{C. reinhardtii} cell are depicted on the right of the figure in the corresponding color. The open red squares depict the speed where the mean shape and stroke both vary with viscosity and are the same as the blue dots in Figure \ref{fig:speed_data_model_compare}.}
    \label{fig:fixed_stroke}
\end{figure}

We first look at the effect of time-varying stroke on the swimming speed by holding the mean shape fixed and varying the stroke with viscosity. Figure \ref{fig:fixed_mean} depicts two examples, where one mean shape is fixed from a cell in viscosity 1.00 cP and the other mean shape is fixed from a cell in viscosity 6.00 cP. When the mean shape is fixed, the changes in speed are small. Since the $R^2$ values are less than $0.14$, this small decrease in speed is not correlated to the change in viscosity. Thus, the time-varying stroke does not have a strong effect on the swimming speed. The changes to swimming speed with varying stroke are on scale with cell variation, as demonstrated in the difference in size of swimming speeds for the different cells in viscosity 1.00 cP in Figure \ref{fig:speed_data_model_compare}.

Figure \ref{fig:fixed_stroke} shows how the mean shape affects the swimming speed by holding the stroke fixed from a specific cell and varying the mean shape across all cells in the data set in the swimming simulations. One fixed stroke is from a cell swimming in water and the other stroke is from a cell swimming in a fluid with a viscosity 6.00 cP. The swimming speeds are similar in Figure \ref{fig:fixed_stroke} whether the time-varying stroke changes or is fixed with viscosity. The data points fall on nearby trend lines with similar slopes and $R^2$ values. The $R^2$ values demonstrate that the correlation between speed and viscosity does not change whether the stroke is fixed or not, which highlights the important effect of the mean shape on the swimming speed.

\begin{figure}[ht!]
    \centering
    \includegraphics[width=.9\linewidth]{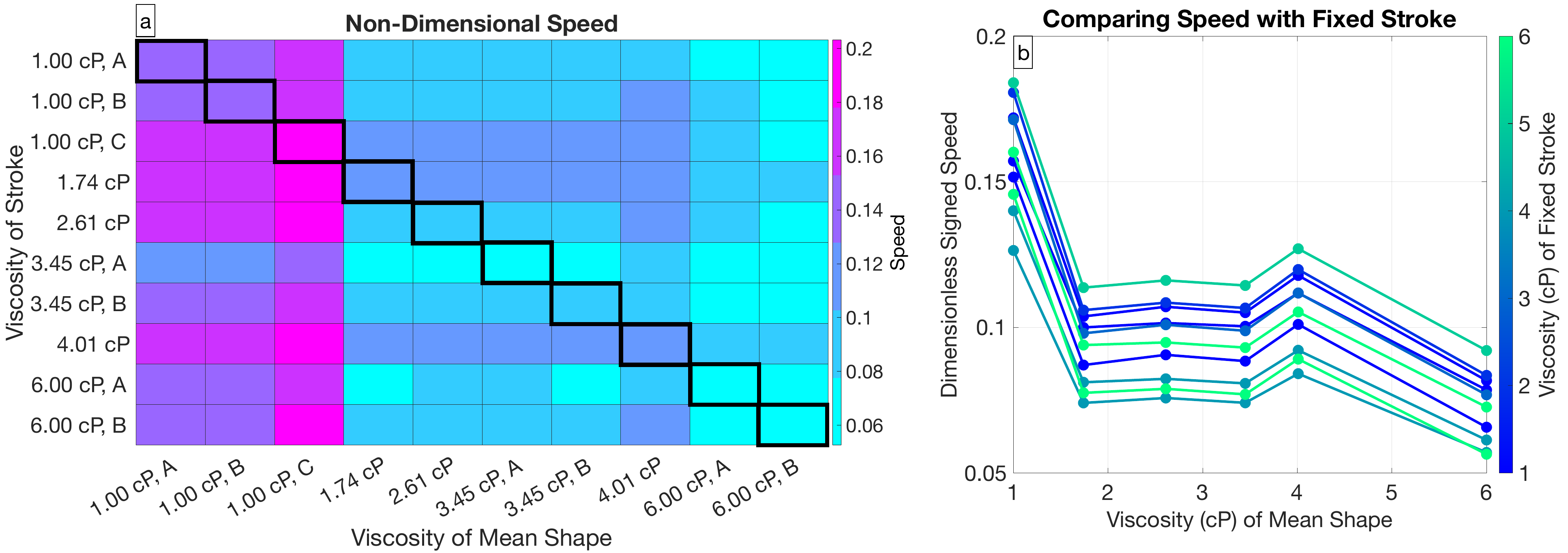}
    \caption{(a) Swimming speed values for each pair of strokes and mean shapes across the data set. Columns corresponds to a fixed mean shape from a specific cell and rows corresponds to a fixed stroke from a specific cell. (b) Comparison of swimming speed with the time-varying stroke fixed and mean shape varying across different cells in different viscosity fluids. When multiple cells in the dataset are in the same viscosity fluid, the average of the different speeds is shown. Each connected line segment represents a simulation with varying mean shape and fixed stroke from a cell in a specific viscosity fluid.}
    \label{fig:mean_speed_heat_map}
\end{figure}

Figures \ref{fig:fixed_mean} and \ref{fig:fixed_stroke} show the effects of mean shape and time-varying stroke on swimming speed for two specific cells. Figure \ref{fig:mean_speed_heat_map}a shows the swimming speed for all combinations of mean shape and time-varying stroke across each cell in the data. Figure \ref{fig:mean_speed_heat_map}b shows that for a fixed stroke, as the mean shape changes, the speed decreases approximately by a factor of two. Note that in this figure, for cells in the same viscosity, we show the average speed across these cells. We further see that the time-varying stroke is not correlated with viscosity by looking at the ordering of the height of the line segments in  Figure \ref{fig:mean_speed_heat_map}b, which do not correspond with viscosity. This result from Figure \ref{fig:mean_speed_heat_map}b is consistent with what was observed in Figure \ref{fig:fixed_stroke}.

\begin{table}[hb]
    \centering
    \begin{tabular}{|c|c|c|c|c|c|c|c|c|c|c|}
    \hline 
       Vis. (cP) & 1.00 & 1.00 & 1.00 & 1.74 & 2.61 & 3.45 & 3.45 & 4.01 & 6.00 & 6.00 \\  \hline 
       \% Var. & 12.30 & 12.62 & 10.58 & 12.41 & 13.45 & 18.08 & 17.00 & 11.21 & 20.59 & 17.02 \\
       \hline 
    \end{tabular}
    \caption{Variation of swimming speed for varying strokes across different cells in different viscosity fluids, calculated by taking the standard deviation of all of the swimming speeds with a particular fixed mean shape and varying stroke and dividing by the "natural" speed, where the stroke and mean shape are from the same cell. The viscosity in the first row denotes that of the fluid from the cell which contributes the mean shape.}
    \label{tab:var_speed}
\end{table}

In Table \ref{tab:var_speed}, for each cell, we compare the variation of the stroke changes where the mean shape is fixed. This table looks at the standard deviation normalized by the "natural" speed, or the speed where the stroke and mean shape are from the same cell. These variations are compared to that of cell variation for swimming speeds from cells with the same viscosity. The variation in swimming speed for cells with viscosity 1.00 cP is 12.84\%, and for cells with viscosity 6.00 cP is 9.59\%. While a few of the values in Table \ref{tab:var_speed} are slightly larger than these cell variation values, the average of the values in this table is 14.53\%, which implies that the majority of minute differences in swimming speed from stroke changes are similar to differences in swimming speed from cell variation. This supports the idea that the changes in stroke are not strongly correlated with changes in viscosity.

Both Figure \ref{fig:mean_speed_heat_map} and Table \ref{tab:var_speed} demonstrate that across all cells, the change in mean shape is the primary cause in the change of non-dimensional speed with viscosity, and the change in time-dependent stroke is not correlated with viscosity changes.

\section{Discussion}
Swimming microorganisms change their gait in response to changes in fluid rheology. We study how strokes change with incremental changes in viscosity over a range between 1 and 6 cP using shape mode analysis and a swimming simulation. We first examine how the dimensionality of the flagellar gait affects the swimming speed. Previously, the shape space has been commonly shown to be two dimensional \cite{werner}, since the first two shape modes make up over 90\% of the data variability. We quantify the effect of adding more shape modes and Fourier modes by examining how the change in reconstruction affects swimming behavior, as seen in Figure \ref{fig:velocity_compare}. We see that with the higher number of shape modes, more frequencies in the Fourier series fit of the time-dependent coefficients are needed. Thus, we use a reconstruction of the flagellar gait involving four shape basis modes and two temporal frequencies in our swimming simulation to quantify how flagellar shape changes affect swimming speed.

The mean shape changes substantially with viscosity, as seen in Figure \ref{fig:mean}, and is shown to be an important contributor to the change in non-dimensional swimming speed. As depicted in Figures \ref{fig:fixed_mean}-\ref{fig:mean_speed_heat_map}, the change in mean shape with viscosity majorly contributes to the change in swimming speed. While the small changes in time-varying stroke from shape mode analysis indicate a small dependence on viscosity, as visualized in Figure \ref{fig:phase_space}, we measure the significance of these changes in stroke by quantifying their effect on the swimming speed, and we find these speed changes to be on the scale of cell variation. Wilson et al. \cite{Wilson_Gonzalez_Dutcher_Bayly_2015} noted that the amplitude of the flagellar gait changes with viscosity. We find that amplitude of the time-varying stroke is not sensitive to fluid viscosity, and the change in mean shape is responsible for the observed change in the amplitude of the flagellar beat.

Different potential sources of a force imbalance inside the axoneme that may produce the time-independent mean shape have been proposed \cite{GEYER20161098, Woodhams}. The changes in mean shape we observe may result from passive deformations of the flagellum from the increased external load from swimming in a fluid of elevated viscosity or through feedback on the mechanochemical origin of the force imbalance. Different mechanochemical feedback mechanisms on motor activity regulating the dynamic flagellar beat have been proposed and analyzed, including tangential deformations (sliding control) \cite{camalet, riedel-kruse}, curvature control \cite{brokaw, hines-blum, Sartori_Geyer_Scholich_Jülicher_Howard_2016}, and normal forces (geometric clutch) \cite{bayly2014, Lindemann_1994}. Alternatively, the dynamic beat may arise from a 'flutter' instability, occurring under axial steady loading \cite{bayly_dutcher, Woodhams}. Previously, Geyer et al. \cite{GEYER20161098} documented the increase in curvature of the mean flagellar shape with changes in ATP concentration and that for a sufficiently large ATP concentration, the amplitude of the time-varying stroke is insensitive to ATP changes. These phenomena are similar to the changes we observe with mean shape and time-varying stroke upon small changes in fluid viscosity. The different responses of the mean shape and time-varying stroke to both fluid viscosity and ATP concentration \cite{GEYER20161098} support the idea that the mean shape and time-varying stroke have independent regulation mechanisms.

\begin{appendices}
\section{Experimental Methods}\label{appendix:experiment}
Experiments with \textit{Chlamydomonas reinhardtii} were performed in Newtonian fluids, which were prepared by adding  Ficoll (Sigma-Aldrich) to M1 buffer solution. The Ficoll concentration varied between 5\% and 20\% by weight in the buffer solution to produce fluids with viscosities ranging between 1 and 6 cP. Fluid viscosity was measured using a stress-controlled rheometer (TA Instruments). Motile \textit{C. reinhardtii} alga were suspended in these Newtonian fluids, where a small volume was stretched to form a thin film (thickness $\approx 20 \mu m$) with a wire-frame device. The cell motion in the thin film was captured using an optical microscope and a high-speed camera. While \textit{C. reinhardtii} flagellar waveform has been shown to be three-dimensional \cite{RufferNultsch, CorteseWan}, the beating of \textit{C. reinhardtii} flagella captured in these experiments is mostly planar due to the experimental setup and confirmation through measurement of the flagellum length and cell body rotation. Any experimental data where the flagellum length deviated more than 10\% or where the cell body rotated significantly was not included. More experimental details can be found in Qin et. al \cite{qin}.

\section{Phase Space Comparison - Ratio Table}\label{appendix:phase_space}
We quantify the apparent change in aspect ratio of the loops in Figure \ref{fig:phase_space} with the number of frequencies used in the reconstruction at different viscosities. Table \ref{tab:ratio} gives the ratios of the two-norm of the Fourier coefficients of $V_{1,N}^j$ and $V_{2,N}^j$ for each cell. The norm of $V_{k,N}^j$ is given as follows, using the definition of $V_{k,N}^j$ from equation \eqref{eqn:v_fit}:
\begin{equation}
    C_{k,N}^j = || V_{k,N}^j(t) ||_2 = \sqrt{ \sum_{n=1}^N \left(A_{k,n}^j\right)^2 + \left(B_{k,n}^j\right)^2}.
\end{equation}
The ratios as shown in Table \ref{tab:ratio} are defined as
\begin{equation}\label{eq:ratio}
    R_N^j = \frac{C_{1,N}^j}{C_{2,N}^j}.
\end{equation}

\begin{table}[ht!]
    \centering
    \begin{tabular}{|c|c|c|c|c|c|c|c|c|c|c|}
    \hline 
         Vis. (cP) & 1.00 & 1.00 & 1.00 & 1.74 & 2.61 & 3.45 & 3.45 & 4.01 & 6.00 & 6.00 \\ 
         \hline 
         \hline 
         $N=1$ & 0.94& 0.98& 0.89& 0.96& 0.99& 1.14&  1.09& 0.99&  1.11& 1.02\\ 
         \hline 
         $N=2$ &     0.91&  0.95&  0.89& 0.96&  0.99& 1.15& 1.10&  1.01& 1.11& 1.07\\ 
         \hline 
         $N=3$ &     0.92& 0.97&  0.89& 0.97& 0.99& 1.15& 1.09& 1.02& 1.12& 1.07\\ 
         \hline 
    \end{tabular}
    \caption{This table describes ratios $R_N^j$, where the ratio is described in equation \eqref{eq:ratio}, for $N = 1,2,3$.}
    \label{tab:ratio}
\end{table}

The first row of the table corresponds to Figure \ref{fig:phase_space}a, the second row corresponds to Figure \ref{fig:phase_space}b, and the last row corresponds to Figure \ref{fig:phase_space}c. Table \ref{tab:ratio} shows that the ratio values are less than 1 for viscosities less than 3 cP, and for viscosities larger than 3 cP, the ratio value is generally larger than 1. This trend in ratio values corresponds to the viscosity trend that was noted in Figure \ref{fig:phase_space} for higher number of frequencies, as the loops of lower viscosity tended to elongate in the $V_2$ direction, which corresponds to the ratios being less than one, and vice versa. However, for one frequency, the ratio trends are visible in the table but not visually apparent in Figure \ref{fig:phase_space}A. Since the quantitative values of the ratios in the table do not change significantly with frequency, they do not provide much information about how the number of frequencies affects the flagellum shape or the swimming behavior.

\section{Simulation Details}\label{appendix:sim_details}
We discretize a model flagellum into equally spaced points, where the spacing is $\Delta s = 0.0833$. The model body is discretized using maximum determinant points on a sphere \cite{Womersley_2020}, with the average spacing between points is approximately $\Delta s$ and then the body is scaled to the shape of an ellipsoid. 

We model the fluid using the steady Stokes equations due to the scale of the problem:
\begin{eqnarray} \label{eq:stokes_momentum}
   -\nabla p +  \mu \Delta \mbf{u} &=& -\mbf{F} \\ \label{eq:stokes_div}
    \nabla \cdot \mbf{u} &=& 0.
\end{eqnarray}
To solve for the swimming speed of a chlamy, we used the method of regularized Stokeslets \cite{cortez}. Using regularized Stokeslets with a set of discrete points on the microswimmer, generally we relate the forces to the velocities as follows:
\begin{equation}\label{eq:stokeslet}
    M \mbf{F} = \mbf{U_L}.
\end{equation}
$\mbf{U_L}$ is the speed in the lab frame and $M$ is the mobility matrix from the regularized Stokeslet. We used the blob function and regularized Stokeslet from \cite{cortez2}. 

To speed up the computation time, we employ both a course mesh and a fine mesh for the cell body. The unknown forces are represented on the course mesh, but the method of regularized Stokelets is employed on the fine mesh. The forces on the coarse mesh are interpolated to the fine mesh using spherical harmonics, and the resulting velocities computed with the method of regularized Stokelets are projected onto the course mesh, again using spherical harmonics.

The reconstruction of the flagellar shapes discussed in Section \ref{sec:shape_analysis} corresponds to the gait in the body frame. Let $\mbf{X_b}$ denote the position of the swimmer in the body frame. To find the position of the swimmer in the lab frame $\mbf{X_L}$, we relate these two positions through a translation and rotation
\begin{equation}\label{eq:lab_frame_position}
    \mbf{X_L} = R (\mbf{X_b} - \mbf{X_0}) + \mbf{X_T},
\end{equation}
where $R$ is a rotation matrix, $\mbf{X_T}$ is the position of a reference point in the lab frame, and $\mbf{X_0}$ is the reference point in the body frame. $\mbf{X_T}$ and $R$ evolve with time in the simulation, and $\mbf{X_T}$ is initially set to the origin while $R$ is initially set to the identity. 

In order to find the swimming speed in the lab frame, we take the derivative of the equation \eqref{eq:lab_frame_position}, and after rearranging, get the following equation: 
\begin{equation} \label{eq:lab_frame_velocity}
    \mbf{U_L} = \mbf{\Omega} \times (\mbf{X_b} - \mbf{X_0}) + R \mbf{U_b} + \mbf{U_T}, 
\end{equation}
where $\mbf{U_b} = \partial_t \mbf{X_b}$ is defined as the body frame velocity. There are three unknowns in this equation: $\mbf{U_L}$, the velocity of the swimmer in the lab frame; $\Omega$, the rotational velocity; and $\mbf{U_T}$, the velocity of the reference point $\mbf{X_T}$. Since we have three unknowns, we add in constraints that the net force and net torque are zero. Using equation \eqref{eq:stokeslet} leads to the following set of equations:
\begin{eqnarray}\label{eq:cons_solve1}
    M\mbf{F} - \mbf{\Omega} \times (\mbf{X_L} - \mbf{X_T}) -  \mbf{U_T} &=& R \mbf{U_b} \\ \label{eq:cons_solve2}
    \sum \mbf{F} &=& \mbf{0} \\ \label{eq:cons_solve3}
    \sum (\mbf{X_L} - \mbf{X_T}) \times \mbf{F} &=& \mbf{0}.
\end{eqnarray}

\end{appendices}

\bibliographystyle{siam}
\bibliography{sample}

\end{document}